\documentclass[10pt,conference]{IEEEtran}

\pdfpageattr{/Group << /S /Transparency /I true /CS /DeviceRGB >>}

\usepackage[utf8]{inputenc}
\usepackage[T1]{fontenc}
\usepackage{textcomp}
\usepackage{fixltx2e}
\usepackage{microtype}
\usepackage{calc}
\usepackage[normalem]{ulem}

\usepackage{lipsum}

\usepackage[range-phrase=--,per-mode=symbol-or-fraction,binary-units=true,range-units=single,list-units=single,detect-all]{siunitx}
\usepackage{silence}
\AtBeginDocument{
	\DeclareSIUnit\bit{bit}
	\DeclareSIUnit\byte{Byte}
	\DeclareSIUnit\decibeli{dBi}
	\DeclareSIUnit\decibelm{dBm}
	\DeclareSIUnit\mph{mph}
	\DeclareSIUnit\resourceblock{RB}
	\DeclareSIUnit\vehicle{veh}
	\DeclareSIUnit\watthour{Wh}
}

\usepackage{csquotes}
\usepackage[backend=biber,style=ieee,doi=false,isbn=false,maxnames=99,mincitenames=1,maxcitenames=2]{biblatex}
\DeclareFieldFormat{sentencecase}{#1} % never apply sentence casing, even if bibtex field is unprotected
\DeclareFieldFormat{titlecase}{#1} % never apply title casing, even if bibtex field is unprotected
\usepackage{xpatch}
\xpatchbibmacro{textcite}{\addspace}{\addnbspace}{}{}
\xpatchbibmacro{Textcite}{\addspace}{\addnbspace}{}{}
\DefineBibliographyStrings{english}{
	andothers = et~al\adddot\addspace
}

\usepackage{acro}

\DeclareAcronym{AI}{short=AI, long=artificial intelligence}
\DeclareAcronym{MC}{short=MC, long=molecular communication}
\DeclareAcronym{RL}{short=RL, long=reinforcement learning}
\DeclareAcronym{RNN}{short=RNN, long=recurrent neural network}
\DeclareAcronym{NN}{short=NN, long=neural network}
\DeclareAcronym{BiRNN}{short=BiRNN, long=bidirectional recurrent neural network}
\DeclareAcronym{CNN}{short=CNN, long=convolutional neural network}
\DeclareAcronym{ML}{short=ML, long=maximum-likelihood}
\DeclareAcronym{IoBNT}{short = {IoBNT}, long = {internet of bio-nano things}}
\DeclareAcronym{ISI}{short= {ISI}, long = {inter-symbol interference}}
\DeclareAcronym{MIMO}{short = {MIMO}, long = {multiple-input multiple-output}}
\DeclareAcronym{CIR}{short=CIR, long=channel impulse response}
\DeclareAcronym{RC}{short=RC, long=reservoir computing}
\DeclareAcronym{PHY}{short=PHY, long=physical layer}
\DeclareAcronym{STO}{short=STO, long=symbol time offset}
\DeclareAcronym{ROC}{short=ROC, long=receiver operating characteristic}
\DeclareAcronym{MAE}{short=MAE, long=mean absolute error}
\DeclareAcronym{AUC}{short=AUC, long=area under the curve}
\DeclareAcronym{CDF}{short=CDF, long=cumulative density function}
\DeclareAcronym{TPR}{short=TPR, long=true positive rate}
\DeclareAcronym{FPR}{short=FPR, long=false positive rate}
\DeclareAcronym{ESN}{short=ESN, long=echo state network}

\usepackage{amsmath}
\usepackage{amssymb}
\usepackage{amsfonts}
\usepackage{amsthm}

\usepackage{booktabs}
\usepackage{url}

\usepackage[inline]{enumitem}

\usepackage[caption=false,font=footnotesize]{subfig}
\usepackage[pdftex]{graphicx}
\DeclareGraphicsExtensions{.pdf,.png,.jpg,.tikz}
\usepackage{tikz}
\usetikzlibrary{arrows}
\usetikzlibrary{calc}
\usetikzlibrary{chains}
\usetikzlibrary{scopes}

\usepackage[american]{babel}
\usepackage{hyphenat}

\usepackage[]{algorithmic}
\usepackage{algorithm}

\usepackage{multirow}
\usepackage{tabularx}
\usepackage{gensymb}
\usepackage{pgfplots}
\pgfplotsset{compat=newest}
\pgfplotsset{plot coordinates/math parser=false}
\usepackage{xspace}

\usepackage[hidelinks]{hyperref}
\usepackage[capitalize]{cleveref}
\Crefformat{figure}{#2Fig.~#1#3}
\Crefformat{section}{#2Sec.~#1#3}
\Crefformat{equation}{#2Eq.~(#1)#3}
\RequirePackage{xstring}
\RequirePackage{xparse}
\RequirePackage[]{acro}
\NewDocumentCommand\acrodef{mO{#1}mG{}}{\DeclareAcronym{#1}{short={#2}, long={#3}, #4}}
\acrodef{ACI}{adjacent channel interference}
\acrodef{AWGN}{additive white Gaussian noise}
\acrodef{BER}{bit error probability}
\acrodef{CSMA}{carrier-sense multiple access}
\acrodef{MAC}{medium access control}
\acrodef{RSS}{received signal strength}

\usepackage{todonotes}

\usepackage[color]{changebar}
\def\todoCtd#1{%
	TODO: #1%
\ifx&#1&...\fi%
	\endgroup
	\cbend
	\relax
}

\def\BibTeX{{\rm B\kern-.05em{\sc i\kern-.025em b}\kern-.08em
    T\kern-.1667em\lower.7ex\hbox{E}\kern-.125emX}}

\usepackage{pdflscape}
\usepackage{tabularx}
\usepackage{enumitem}
\usepackage{array}
\usepackage{comment}
\usepackage[subtle]{savetrees}

\newcolumntype{P}[1]{>{\centering\arraybackslash}p{#1}}

\usepackage[
  range-phrase=\text{ to },
  per-mode=symbol-or-fraction,
  range-units=single,
  list-units=single,
  detect-all,
  list-final-separator={, },
]{siunitx}
\DeclareSIUnit{\molar}{M}

\begin{document}

\title{Learning When to Listen: Physics-Informed Synchronization in Molecular Channels}

\author{
    \IEEEauthorblockN{
        Osman Tugay Basaran\IEEEauthorrefmark{1}\IEEEauthorrefmark{2}, 
         Saswati Pal\IEEEauthorrefmark{1}, 
        Falko Dressler\IEEEauthorrefmark{1},
        Jorge Torres G\'omez\IEEEauthorrefmark{1}
    }
    \IEEEauthorblockA{
        \IEEEauthorrefmark{1}School of Electrical Engineering and Computer Science, TU Berlin, Germany \\
        \IEEEauthorrefmark{2}Optical Zeitgeist Laboratory, INRS, Montréal, QC, Canada\\
         \{basaran, pal, dressler, torres-gomez\}@ccs-labs.org
        }
}

\maketitle

\begin{abstract}
Recent advances in chemical computing create new opportunities to realize biocompatible communication devices, which devise various application in precision medicine.
Implanted microscopic transmitters and receivers can exchange information using molecules that diffuse through a biological medium, enabling them to form a network and jointly coordinate sensing and actuation operations.
However, randomness in molecular propagation causes severe signal distortion, and receivers have limited computational capacity to compensate.
The receiver design must therefore combine robustness to diffusion-induced distortion with computational architectures amenable to implementation in the chemical domain.
To meet these requirements, we propose a physics-informed \ac{RL} framework implemented with the \ac{RC} architecture.
The physics component of the architecture provides the robustness needed in \ac{MC} channels, while the \ac{RC} architecture exploits the computational capabilities already presente in free-diffusion channels.
In this contribution, we apply this framework to the task of synchronizing the receiver's clock with a sequence transmitted through an \ac{MC} channel.
The physics part approximates the maximum-likelihood function synchronization objective, while the other learns a timing policy using this objective as its reward.
Simulation results yield clock misalignments above half a second, compatible with molecular transmissions operating on the timescale of seconds.
\end{abstract}


\section{Introduction}

New nanoscale technologies are under development for precision medicine applications, where networks of biocompatible sensors are envisioned to operate collaboratively inside the human body, detecting early signs of diseases~\cite{lagasse2023future}.
This vision challenges their communication capabilities, prompting the research community to engineer \ac{MC} links to enable their connectivity~\cite{soldner2020survey}.
However, the nature of \ac{MC} channel poses new challenges for establishing a reliable communication link.
In diffusion-based \ac{MC} channels (like in the extracellular matrix), molecules released by the emitter start diffusing to the receiver, where random molecular arrivals and receptor-binding events give rise to temporal dynamics and observation uncertainty~\cite{jamali2019channel}.
Motivated by the need for robust yet chemically realizable receivers, we develop a physics-informed \ac{RL} synchronization framework that integrates \ac{RC} modules as implemented in the molecular domain.

The research community is increasingly incorporating \ac{AI} models to enable communication over \ac{MC} channels~\cite{torres-gomez2026communicating}.
Learning-based architectures have been investigated to implement basic communication components such as channel estimation, synchronization, and symbol detection~\cite{ozdemir2024estimating, debus2024synchronized, torres-gomez2023explainability}.
However, most existing models are resource-intensive and are often implemented as generic black-box networks.
A closer look at the reported hyperparameters reveals models with hundreds of learnable coefficients~\cite[Tab.~II]{torres-gomez2026communicating}, which limits their deployment in the reality of microdevice solutions operating in \ac{MC} channels.
Besides, black-box models do not explicitly preserve the underlying dynamics of the \ac{MC} channel~\cite{goring2024out-of-domain} and may therefore fail to cope with stochastic channel variations beyond those represented during training.
Explicitly accounting for \ac{MC} channel dynamics within the learning architecture is particularly consequential because receiver-side learning must satisfy two coupled design requirements. 
For an efficient operation, the \ac{AI} model must remain faithful to the stochastic physics of molecular propagation and reception, and it must operate within the limited computational resources of biocompatible receivers. 
This motivates new designs of learning architectures that combine physical faithfulness, and low trainable complexity.

To avoid high resource demands on the \ac{AI} model, we introduce a physics-based \ac{RL} framework for a synchronization task.
The physics part accounts for the maximum likelihood estimator and provides a reward to the \ac{RL} agent, which is trained to adjust the clock offset at the receiver.
Furthermore, we develop the physics and the \ac{RL}-agent using a \ac{RC} framework, which has been recently developed utilizing diffusion channels as reservoirs~\cite{bilge2026reservoir,uzun2026molecular}.
Exploiting the reservoir dynamics of diffusion channels is particularly convenient as it restricts training to lightweight readout layers, reducing complexity compared with fully trainable recurrent receivers.
Following this line of developments, our key contributions are summarized as follows:
\begin{itemize}
    \item We formulate symbol synchronization in \ac{MC} as a {physics-informed \ac{RL} problem}, in which the analytical maximum-likelihood objective is retained as a reward.
    \item We develop a {dual-\ac{RC} architecture} that separates likelihood approximation from timing-policy learning.
    The physics module learns a surrogate of the maximum-likelihood objective, while the \ac{RL}-agent develops the policy.
    \item We validate the proposed architecture to show that {\ac{RC}-PHY preserves the overall behavior of the exact physical likelihood, while the complete \ac{RC}-\ac{RL} architecture attains a \ac{STO} above half a second.}
\end{itemize}

\section{State of the Art}
\label{sec_sota}

Recent learning-based \ac{MC} research has introduced neural models at different stages of the receiver chain~\cite{torres-gomez2026communicating}. 
Channel-surrogate models estimate received molecule fractions in complex three-dimensional geometries and \ac{MIMO} channels~\cite{ozdemir2024estimating,cheng2024channel}. 
These models reduce reliance on closed-form channel expressions and repeated particle simulations. 
However, the learned surrogate must approximate the complete mapping from channel parameters to receiver observations for subsequent detection or synchronization.
Moreover, such surrogates primarily predict expected channel parameters rather than preserving the sample-level stochastic structure required for likelihood-based inference.

An alternative is to learn the receiver mapping directly from molecular observation sequences. 
Informer-based and recurrent architectures exploit temporal dependencies for symbol detection under unknown or time-varying channel conditions~\cite{cheng2024informer,xiang2025hybrid}. 
Convolutional and gated recurrent networks have further been used for joint synchronization and non-coherent detection~\cite{casaleiro2024synchronization}. 
These results show that temporal models can infer timing and symbol information directly from received molecular observations, even when the instantaneous channel state is not explicitly available. 
%However, the physical channel model is mainly used to generate the training data; once trained, the network makes receiver decisions without explicitly evaluating the underlying stochastic reception model.

When the physical reception model is not retained in the inference rule, the learned receiver is not explicitly constrained by the underlying stochastic dynamics. 
Physics \ac{AI} addresses this limitation by retaining known physics and learning only unresolved components. 
Such structural information becomes particularly important beyond the training regime, where unconstrained dynamical models may fail~\cite{goring2024out-of-domain}. 
Hybrid models further provide a practical realization of this idea by coupling a mechanistic solver with a learned correction~\cite{belbute2020combining}.
For \ac{MC}, the stochastic reception likelihood provides a natural physical anchor for inference~\cite{heinlein2023stochastic}. 
Learning can then be restricted to dynamics that are not adequately captured by the nominal channel model. 
% Physics-preserving correction models have adopted the broader principle of retaining a mechanistic model while learning its unresolved correction~\cite{wang2024p2c2net}.
However, these methods target physical-state prediction rather than stochastic receiver inference and do not address the resource constraints of \ac{IoBNT} receivers.

Addressing these constraints requires the learned component itself to remain lightweight.
Reservoir computing achieves this by fixing the recurrent dynamics and training only the readout~\cite{dong2020reservoir}. 
In \ac{MC}, this architecture has recently been applied to time-varying \ac{ISI} detection~\cite{bilge2026reservoir}, but it still does not incorporate the physics of the communication process into the architecture. 
%However, the reservoir acts as the decision model rather than as a correction to an explicit stochastic reception model.
Taken together, existing approaches leave open how to adapt to unresolved temporal dynamics without replacing the stochastic likelihood-based inference rule or relying on a fully trained recurrent model. 
We address this problem by retaining the physical reception likelihood and restricting learning to a low-dimensional temporal correction, which is then used for synchronization.

\section{System Model and Problem Description}
\label{sec_problem}

\begin{figure}
    \centering
  \includegraphics[width=\columnwidth]{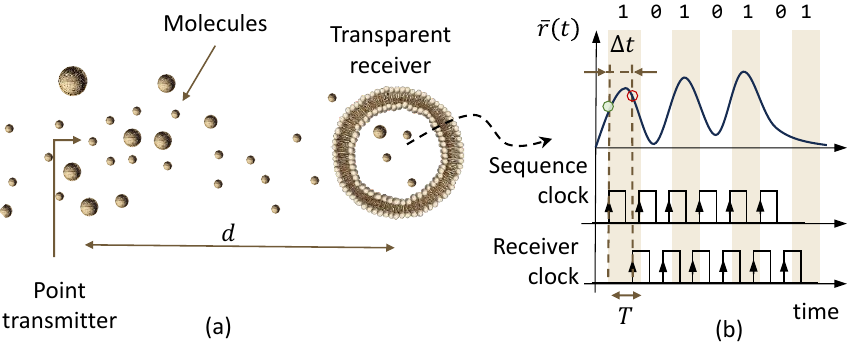}
  \caption{Illustration of the system model.}
  \label{fig_synchronization}
\vspace{-1em}
\end{figure}

We study a molecular communication link where molecules that freely diffuse between a point transmitter and a spherical transparent receiver are the carriers of information; see an illustration in \Cref{fig_synchronization}~(a).
Bits are carried in the concentration of molecules, where emitting a one corresponds to the point transmitter releasing an impulse of $M$ molecules into the medium, and emitting a zero releases no molecules, with emissions performed at clock period $T$.
The receiver is assumed to be a transparent spherical receptor of volume $V_r$, which is placed at a distance $d$ from the point transmitter.
In this diffusive medium, the number of received molecules follows a Poisson distribution of parameter given by their average number as~\cite[Eq. (74)]{jamali2019channel}
\begin{align}\label{eq_channel}
    r(t)&=\mathcal{P}(\overline{r}(t-\Delta t)),\quad
    \overline{r}(t)=\sum_{l=0}^{L-1}{a_lh(t-lT)},\\
    h(t)&=\frac{V_r}{(4\pi D t)^{3/2}}e^{\frac{-d^2}{4Dt}}\nonumber
\end{align}
where $\bar{r}(t)$ denotes the average number of received molecules, see a representation in~\Cref{fig_synchronization}~(b), $a_l=\{0,M\}$ denotes the transmitted binary sequence, $M$ is number of molecules released for emitting a one, $h(t)$ is the \ac{CIR} of the diffusive channel, $L$ is the channel memory, $D$ is the diffusion coefficient of the released molecules~\cite[Eq.~(34)]{jamali2019channel}, and $V_r$ is the volume of the transparent receiver.

To decode the sequence of ones and zeros properly, the receiver must first acquire the symbol timing from $r(t)$. 
The receiver's clock must align with the received sequence by estimating the clock offset $\Delta t\in[0,T)$ as illustrated in \Cref{fig_synchronization}~(b).\footnote{Here, $\Delta t$ refers to the time difference between the received symbol boundaries and the rising edges of the receiver clock, see a representation in~\Cref{fig_synchronization}~(b).}
This time difference depends on the time needed for the particles to travel from the point transmitter to the receiver, which is random in nature (see~\Cref{eq_channel}) and also depends on unknown parameters like the distance $d$ to the emitter via $h(t)$, assuming $D$ and $V_r$ are already specified.
In this setup, acquiring the symbol timing is referred to as the synchronization problem, and it is optimally addressed by applying the maximum--likelihood formulation as~\cite[Eqs. (4)--(5)]{jamali2017symbol}
\begin{equation}\label{eq_problem}
\Delta \tilde{t}=\underset{\Delta t \in [0,\,T)}{\mathrm {argmax}}\text{ } \ln\Lambda(\Delta t|\textbf{y}),\text{ } \Lambda(\Delta t|\textbf{y})=\prod _{n=1}^N \mathcal {P}(\bar{r}(t_{n}-\Delta t)),
\end{equation}
where $\Lambda(\Delta t|\textbf{y})$ is the likelihood function.
This function takes as observation the sequence $r(t_n)$ along the observation window $T$ represented with
\begin{equation}\label{eq_y}
\mathbf{y}=[r(t_1),\ldots,r(t_N)]^{\top},
\end{equation}
where $N=\lfloor T/t_s\rfloor$ is the number of samples within the observation window of duration $T$, $t_s$ is the sampling time, $t_{n}=(k-1)T+(n-1)\,t_s$ denotes the sampling sequence of time index $n\in\mathbb{N}$, and $k\in\mathbb{N}$ indexes the window segment where the evaluation is carried out.
Besides, with the definition of the Poisson distribution, the \Cref{eq_problem} can be simplified to
\begin{equation}\label{eq_ML}
\ln\Lambda(\Delta t|\textbf{y})=\sum_{n=1}^{N}
\left[r(t_n)\text{ }\ln\,\bar{r}(t_n-\Delta t)\right],
\end{equation}
after omitting the terms $\bar{r}(t_n-\Delta t)$ and $\ln{\bar{r}(t_n-\Delta t)!}$ from the full expansion of $\ln\Lambda(\Delta t|\textbf{y})$ as given in~\cite[Eq. (18)]{jamali2017symbol}.
This linearity in the observation vector $\mathbf{y}$ makes the likelihood function particularly suitable for approximation using a linear readout, a condition we exploit in the next section via the \ac{RC} architecture.

% ============================================================
\section{Dual RC-architecture}
\label{sec_methodology}
% ============================================================
    
\begin{figure}
        \centering
        \includegraphics[width=\linewidth]{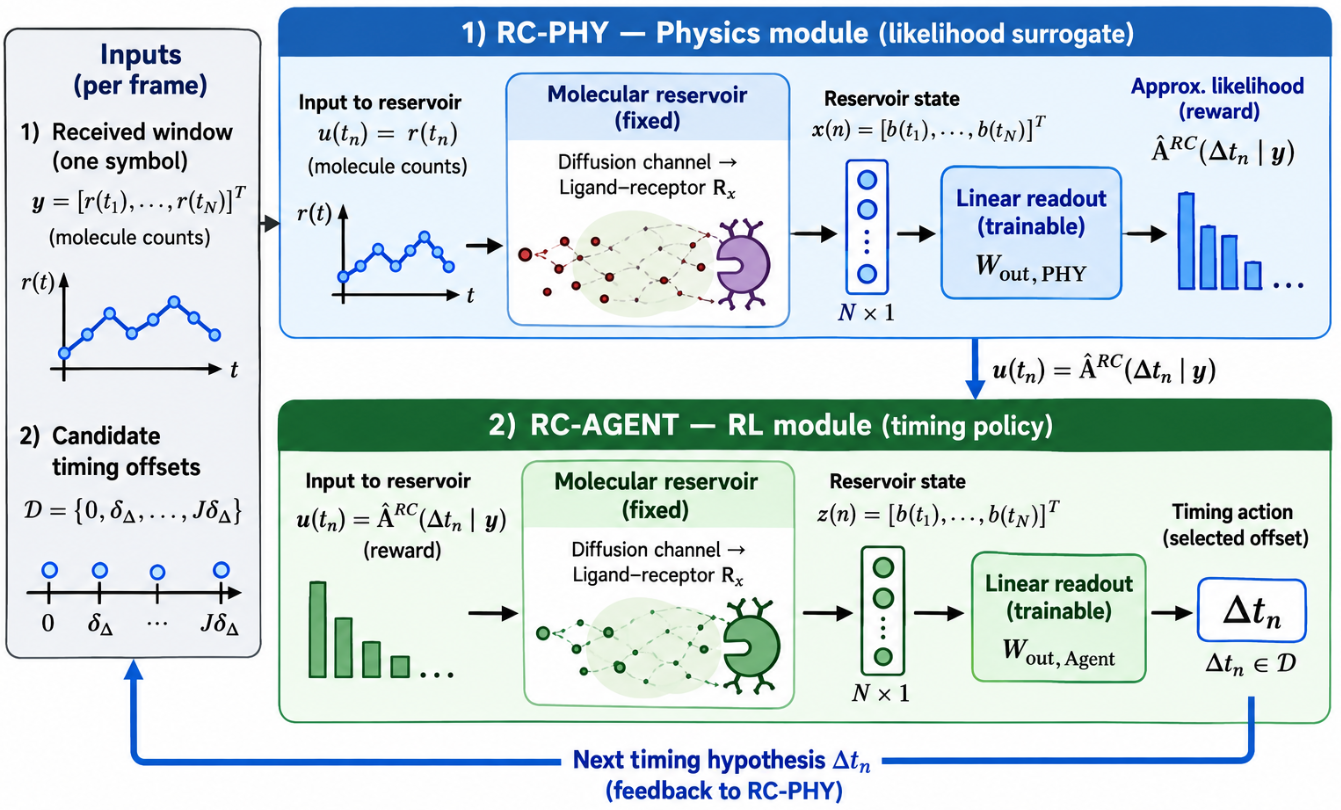}
        \caption{Overview of the physics-informed \ac{RC}-RL synchronization framework.}
        \label{fig_architecture}
        \vspace{-1em}
\end{figure}
    
We use the \ac{RC} architecture to assemble a physics-informed \ac{RL} framework, where the physics component directly evaluates the maximum--likelihood synchronization objective in \Cref{eq_ML}, while the \ac{RL}-agent task is to find the value of $\Delta t$ that maximizes $\ln\Lambda(\Delta t|\textbf{y})$.
We illustrate this framework and its interfaces in \Cref{fig_architecture}, where both components run on separate \ac{RC} architectures.
The physics module implements a surrogate of the maximum--likelihood function in \Cref{eq_ML}, it is fed with the vector $\mathbf{y}$ in \Cref{eq_y} and the estimated $\Delta t$ as performed by the \ac{RL}-agent.
The \ac{RL}-agent uses the maximum-likelihood evaluation as the reward and outputs an estimate of $\Delta t$ as its action.
The agent actions are constrained within a finite space as 
\begin{equation}\label{eq_delta_t}
\Delta t_n\in\mathcal{D}=\{0,\,\delta_{\Delta},\,2\delta_{\Delta},\ldots,(J-1)\delta_{\Delta}\},
\end{equation}
where $\delta_{\Delta}$ as a fixed action-step size, $J=\lfloor\frac{T}{\delta_\Delta}\rfloor$ is the dimension of $\mathcal{D}$, and $n$ denotes the time index of the system.
The physics and the agent modules are trained in two steps for separate tasks: first, the physics module is trained to approximate $\ln\Lambda(\Delta t|\textbf{y})$ in~\Cref{eq_ML}; second, the \ac{RL}-agent is trained to maximize the reward given by the likelihood function, and while operating in the loop with the first module.

\subsection{Reservoir computing description and capabilities}
\label{sec_RC}

\begin{figure}
    \centering
  \includegraphics[width=.9\columnwidth]{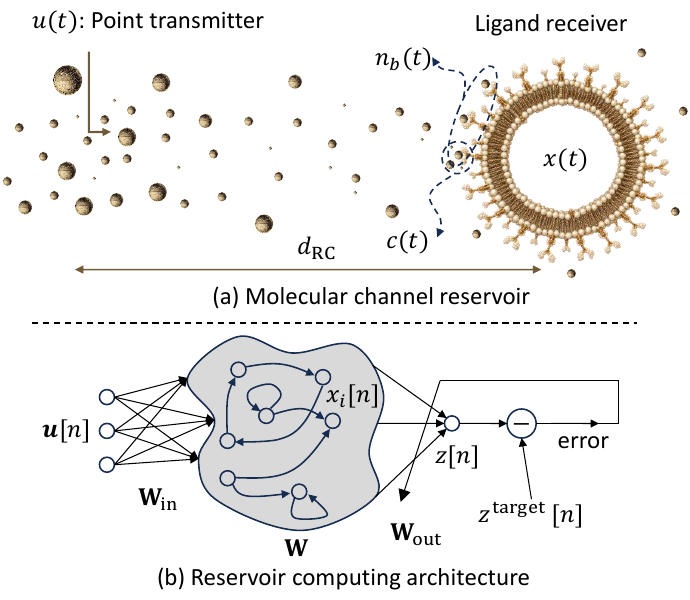}
  \caption{Illustration of the \ac{RC} module.}
  \label{fig_reservoir}
\vspace{-1em}
\end{figure}

Both modules in \Cref{fig_architecture} use an \ac{RC} architecture, which comprises a reservoir and linear readout components.
The linear readout is the only trainable component, eliminating the need to optimize the internal weights, in contrast to conventional \ac{RNN} architectures.
The reservoir component expands the input signal (by echoing it within the reservoir internal nodes) into a domain with enough dimension to allow the linear readout to fulfill the given task.
In our case, the reservoir component is conceived through a point transmitter, a free-diffusion \ac{MC} channel, and a ligand-receptor receiver,  which mimics the design introduced in~\cite[Fig. 1(d)]{uzun2026molecular} and illustrated in \Cref{fig_reservoir}~(a).
The \ac{RC} operates on the observation clock, where its input is presented once per sampling interval, so the reservoir step is the sampling time $t_s$ and the transmitter releases $u(t_n)$ molecules per step.
Released molecules propagate by free diffusion toward the ligand-receptor receiver according to the same \ac{CIR} ($h(t)$ ) as in \Cref{eq_channel}, evaluated at a separation distance $d_{\mathrm{RC}}$ and with $V_r=1$ to approximate the ligand-receptor as a point receiver.
The resulting concentration is $c(t_n)=\sum_{l=-\infty}^\infty u(t_{n-l})\,h(t_l)$, with $t_l=l\,t_s$, $t_n=n\,t_s$, $n,l\in\mathbb{Z}$, where $L_\mathrm{RC}$ is the channel memory of the \ac{RC} link in number of symbols.
At the receiver, the fraction of bound receptors $b(t)=n_b/N_R$ evolves as $db/dt=k_\mathrm{on}c(t)[1-b(t)]-k_\mathrm{off}b(t)$, where $n_b$ is the number of occupied receptors of a total of $N_R$, $k_\mathrm{on}$ and $k_\mathrm{off}$ are the binding and unbinding constants, respectively, and determines the dynamics of the ligand receptor.
In this setup, the echo state is defined as a delay line of the occupancy samples, given as~\cite[Eq. (8)]{uzun2026molecular}
\begin{equation}\label{eq_x}
\mathbf{x}[n]=[b(t_n),b(t_{n-1}),\dots,b(t_{n-N+1})]^\top, 
\end{equation}
which retains the temporal receptor dynamics over one
symbol interval. When fewer than the available observation samples are used,
the receptor trajectory is uniformly subsampled to obtain an
$N$-dimensional virtual-node representation, allowing the readout complexity
to be reduced without changing the underlying molecular dynamics.
The vector $\mathbf{x}[n]$ is used to evaluate the linear readout layer as $z[n]=\mathbf{W}_{\mathrm{out}}^{\top}\mathbf{x}[n]$~\cite[Eq.~(9)]{uzun2026molecular}, in this way completing the \ac{RC} architecture.

The physical reservoir described above can be cast into the equivalent
\ac{ESN} representation illustrated in \Cref{fig_reservoir}~(b), defined by the reservoir and readout layer components as~\cite[Eqs. (1)--(2)]{jaeger2001echo} 
\begin{align}\label{eq_reservoir}
    \text{Reservoir: }\mathbf{x}[n]&=f(\mathbf{W}\mathbf{x}[n-1]+\mathbf{W}_\mathrm{in}\mathbf{u}[n]) \\
    \text{Readout layer: }z[n]&=f_\mathrm{out}(\mathbf{W}_\text{out}^\top \mathbf{x}[n]),\nonumber
\end{align}
where the matrices $\mathbf{W}$ and $\mathbf{W}_\mathrm{in}$ define the activation of the $x_i$'s nodes, $\mathbf{W}_\text{out}^\top$ is the trainable matrix, and the functions $f$ and $f_\mathrm{out}$ are arbitrary defined.
To operate as an \ac{ESN}, the physical reservoir in \Cref{fig_reservoir}~(a) must satisfy the echo-state property~\cite[Def.~1]{jaeger2001echo}.
This condition requires that as $n \to \infty$, the dynamics of each node in $\mathbf{x}[n]$ \Cref{eq_x} become independent of its initial state $\mathbf{x}[0]$. Starting from \Cref{eq_x}, we derive a sufficient condition in \Cref{sec_echo_app} for the physical reservoir to satisfy this property
\begin{equation}\label{eq_cond}
d_{\mathrm{RC}}\; > \;
\frac{k_{\mathrm{on}}\,M_{\mathrm{RC}}}
{4\pi D\,(1-t_sk_{\mathrm{off}})}.
\end{equation}
which can be used to delimit the minimum separation in the reservoir in terms of components fixed by molecular-related parameters ($D,\,k_\mathrm{on},\,k_\mathrm{off},\,M$) and the defined sampling time ($t_s$).

Besides, we model the physical reservoir in \Cref{fig_reservoir}~(a) with a linearized model.
We find the time-independent matrices $\textbf{W}\in\mathbb{R}^{_{N\times N}}$ and $\textbf{W}_\mathrm{in}\in\mathbb{R}^\mathrm{{N\times NL_\mathrm{RC}}}$ that make \Cref{eq_reservoir} approximate the echo state vector in \Cref{eq_x}. 
Departing as well from the physics in \Cref{eq_x}, we derive in the \Cref{sec_W_app} an approximate \ac{ESN} linearized model of the reservoir in \Cref{fig_reservoir}~(b) yielding
\begin{align}\label{eq_matrices}
\mathbf{W}&=\begin{bmatrix}
\bar\alpha & 0 & \cdots & 0 & 0\\
1 & 0 & \cdots & 0 & 0\\
0 & 1 & \cdots & 0 & 0\\
\vdots & & \ddots & & \vdots\\
0 & 0 & \cdots & 1 & 0
\end{bmatrix},\\
\mathbf{W}_\mathrm{in}&=t_sk_\mathrm{on}\begin{bmatrix}
h(t_1) & \cdots & h(t_{NL_\mathrm{RC}})\\
0 & \cdots & 0\\
\vdots & & \vdots\\
0 & \cdots & 0
\end{bmatrix},\nonumber
\end{align}
where $\bar{\alpha}$ is given in \Cref{eq_alpha} in the apppendix, $\bar{c}$ is the average molecule concentration within an $N$-sample time window centered at $t_n$, see \Cref{eq_bar_c}, while $f$ and $f_\mathrm{out}$ in \Cref{eq_reservoir} are the identity functions.
Besides, we evaluated the spectral radius of $\mathbf{W}$ in \Cref{sec_echo_app} to determine the reservoir's capacity to echo the input $u(t_n)$, yielding
\begin{equation}\label{eq_rho}
    \rho(\mathbf{W})=|\bar\alpha|=|1-t_s\big(k_\mathrm{off}+k_\mathrm{on}\bar c\big)|
\end{equation}
Smaller values of $\rho(\mathbf{W})$ correspond to weaker echoing and reduced dynamic richness in the state vector $\mathbf{x}$, whereas larger values increase representation diversity at the expense of stability.
Ideally, $\rho(\mathbf{W})$ should be close to $1$ from below (see~\cite[Fig. 3]{verstraeten2007experimental}), which can be tuned via the distance $d_\mathrm{RC}$ in $\bar c$ for fixed parameters $t_s$, $k_\mathrm{off}$, and $k_\mathrm{on}$.

\subsection{Integrating the RCs within the RL framework}

Two identical separate \ac{RC} modules are integrated for the physics (\ac{RC}--PHY) and the \ac{RL}-Agent components.
They are connected as follows
\begin{align}\label{eq_rc_phy}
&\text{\ac{RC}--PHY: }
u[n]=r(t_n),\text{ } \mathbf{x}[n]=\mathbf{W}\mathbf{x}[n-1]+\mathbf{W}_\mathrm{in}\mathbf{u}[n],\\
&z_\mathrm{PHY}[n]=\mathbf{W}_{\mathrm{out,PHY}}^{\top}\mathbf{x}[n]\text{ with } z^\text{target}_\mathrm{PHY}(t_n)=\ln\Lambda(\Delta t\!=\!p\,t_s\mid\mathbf{y}),\nonumber
\\[0.5em]
&\text{\ac{RC}--Agent: } u[n]=z_\mathrm{PHY}[n],\text{ } \mathbf{x}[n]=\mathbf{W}\mathbf{x}[n-1]+\mathbf{W}_\mathrm{in}\mathbf{u}[n],\nonumber\\
&z(n)=\mathbf{W}_{\mathrm{out,Agent}}^{\top}\mathbf{x},\text{ with } z^\text{target}_\mathrm{Agent}(t_n)=\Delta t,\nonumber
\end{align}
where the matrices $\textbf{W}$ and $\textbf{W}_\mathrm{in}$ are identical for both modules as they are fixed by the same \ac{MC} dynamics in \Cref{fig_reservoir}~(a), and both modules are connected such that the output of \ac{RC}--PHY feds the input of \ac{RC}--Agent module.
Only the readout layer matrices $\mathbf{W}_{\mathrm{out,PHY}}$ and $\mathbf{W}_{\mathrm{out,Agent}}$ are evaluated with the separate training of both modules for the given $z^\text{target}(t_n)$'s as specified above.
The training of both \ac{RC} modules follows the standard ridge-regression formulation in~\cite[Eq.~(27.10)]{lukosevicius2012practical} as
\begin{equation}\label{eq_training}
\mathbf{W}_{\mathrm{out}}=\arg\min_{\mathbf{W}_{\mathrm{out}}}\;\sum_{k=1}^{K}\big(z[k]-z^\text{target}[k]\big)^2+\beta\big\|\mathbf{W}_{\mathrm{out}}\big\|_2^2,
\end{equation}
which is solved numerically.\footnote{We also provide access to the code that implements this architecture under the link \url{ https://github.com/tkn-tub/learningwhentolisten}}

% -------------- Section end marker --------------
%                _       _
%               ( )_    ( )
%    ___  _   _ | ,_)   | |__     __   _ __   __
%  /'___)( ) ( )| |     |  _ `\ /'__`\( '__)/'__`\
% ( (___ | (_) || |_    | | | |(  ___/| |  (  ___/
% `\____)`\___/'`\__)   (_) (_)`\____)(_)  `\____)
%

\section{Simulation Results}
\label{sec_results}

We evaluate the proposed dual-\ac{RC} framework using molecular observations generated from the free-diffusion channel model in \Cref{eq_channel}.
The release sequence is generated with symbol duration $T=\SI{4}{\second}$ and sampling interval $t_s=\SI{0.1}{\second}$, yielding $N=40$ samples per symbol.
The candidate timing space is discretized over $[0,T)$ with resolution $\delta_{\Delta}=\SI{0.1}{\second}$, resulting in $J=40$ candidate timing offsets.
For both components, we use $N=40$ virtual nodes, corresponding to the full
receptor-state representation over one symbol interval. The ridge coefficient
in \Cref{eq_training} is set to $\beta=\num{3e-4}$.
Following these parameters, the resulting average
concentration $\bar c$ determines the spectral radius as $\rho(\mathbf{W})=0.89$, see \Cref{eq_rho}.
This value remains below unity, satisfying the echo-state stability condition while retaining sufficient fading memory for the downstream timing-estimation task.

Molecular observations are generated using a Poisson model with $M=8000$ released molecules per transmitted bit.
To represent a physically realistic nanoscale diffusion scenario, we use the diffusion coefficient for vesicles in interstitial fluid as $D=\SI{3.3}{\micro\meter\squared\per\second}$~\cite[Tab.~II]{zoofaghari2024modeling}, receiver radius as $r_{\mathrm{Rx}}=\SI{4}{\micro\meter}$~\cite[Tab.~III]{zoofaghari2024modeling}, and transmitter--receiver separation as $d=\SI{11}{\micro\meter}$, which is equivalent to a surface-to-surface separation of $\SI{3}{\micro\meter}$ among cells~\cite[Fig.~9]{zoofaghari2024modeling}.
For the \ac{RC} scenario in \Cref{fig_reservoir}~(a), we use the same diffusion coefficient $D$, while the internal reservoir link is parameterized independently from the communication link.
We denote by $M_{\mathrm{RC}}$ the number of molecules used to excite the
physical reservoir and select $d_{\mathrm{RC}}=\SI{3}{\micro\meter}$ such
that the echo-state condition in \Cref{eq_cond} is satisfied while remaining
smaller than $r_{\mathrm{Rx}}$.
The ligand rates are set to $k_{\mathrm{on}}=\SI{6.022e8}{\per\molar\per\second}$ and $k_{\mathrm{off}}=\SI{1}{\per\second}$ as given in \cite[Tab.~I]{uzun2026molecular}.

We compare three synchronization strategies: \textit{Exact-Physics-ML}, which directly maximizes the analytical likelihood in \Cref{eq_ML} and therefore serves as the optimal baseline, \textit{Direct-\ac{RC}--Argmax}, which selects the candidate offset that maximizes the learned \ac{RC}--PHY likelihood, isolating the performance of the physical surrogate without the \ac{RC}--Agent module, and \textit{Threshold} which provides a lightweight, non-learning baseline in which the received molecular signal $r(t_n)$ is compared against a threshold.
We decide a peak is present whenever the output of these synchronization strategies exceeds a given threshold $\theta$.
For a fair comparison, we first normalize each method's output by its maximum, ensuring the same threshold range applies to all synchronization strategies.
We then fix the threshold to the value that maximizes the noise margin and keep it fixed for the subsequent evaluation of the \ac{CDF} of the \ac{STO}, as described next.

\begin{figure}
    \vspace{-.8em}
    \centering
    \subfloat[\ac{RC}--PHY surrogate fidelity.]{%
    \includegraphics[width=0.5\linewidth]{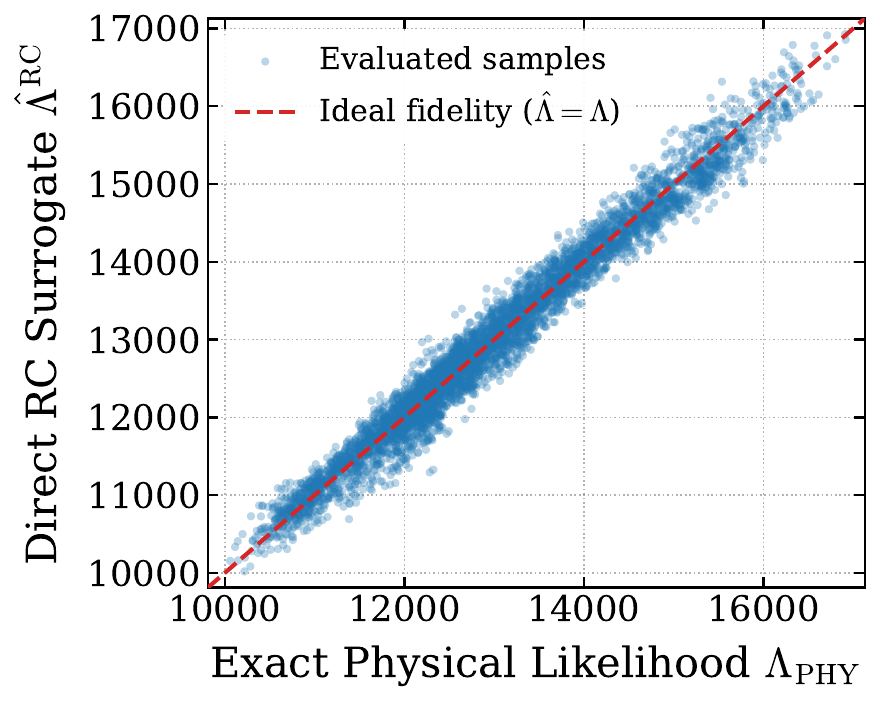}%
    \label{fig_surrogate}%
    }%
        \hfill
    \subfloat[\ac{CDF} of the absolute \ac{STO} error.]{%
        \includegraphics[width=0.5\linewidth]{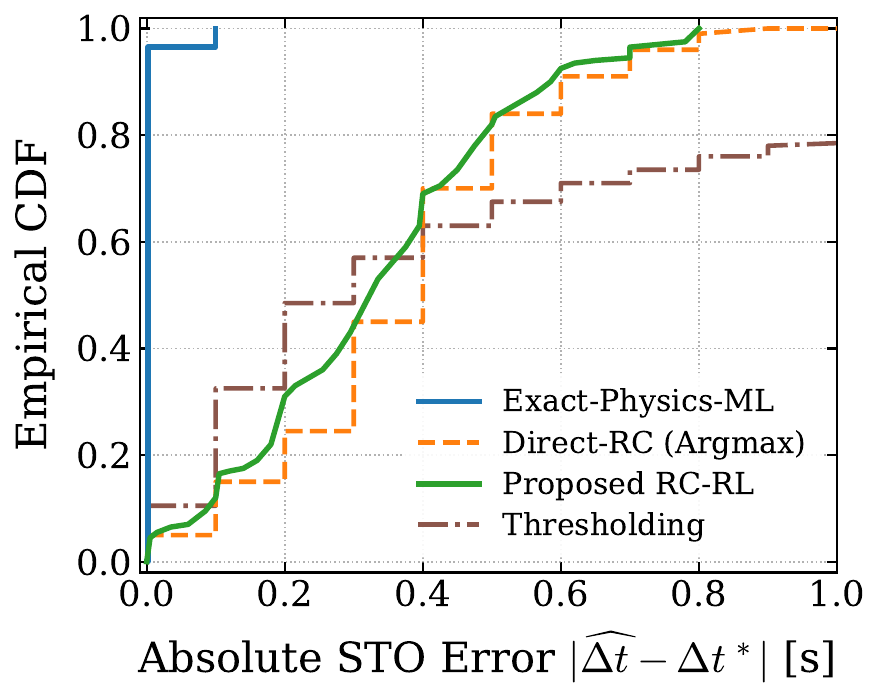}%
        \label{fig_cdf}%
    }%
   \caption{Synchronization performance.}
    \label{fig_xtn}
    \vspace{-1em}
\end{figure}

We evaluate the trained models in \Cref{eq_rc_phy} on the test observation vector in~\Cref{eq_y} and the candidate offsets $\Delta t_n$ in \Cref{eq_delta_t} to assess how accurately \ac{RC}--PHY approximates the maximum--likelihood objective in \Cref{eq_ML}.
As shown in \Cref{fig_surrogate}, the surrogate follows the trend of the exact likelihood function, where the $\SI{90}{\percent}$ of the median deviation corresponds to approximately $\SIrange{2}{4}{\percent}$ of the likelihood magnitude.
These results are sustained by the reservoir's memory capacity, set to $N=40$, which equals the symbol time over which the variability of the input $r(t_n)$ is retained.
Even though the surrogate error can affect the downstream synchronization task, the 	\ \ac{RC}--Agent compensates for it, as shown in the \ac{STO} performance below.

A comparison of the \ac{STO} performance among the synchronization strategies is shown in \Cref{fig_cdf}, which is evaluated as $\text{STO}=|\widehat{\Delta t}-\Delta t|$.
We measure \ac{STO} after fixing the detection threshold with the median between $1$ (since peaks are normalized) and $h(t_\text{peak}+T)$ (which measures the interference level to the next symbol, assuming unit channel memory), and $t_\mathrm{peak}$ is the peak time of the received signal $r(t)$ along a symbol time.
At the $90$th percentile, Exact-Physics-ML attains a \ac{STO} least than $\SI{100}{\milli\second}$ (optimal baseline), while realistic cases exhibits an \ac{STO} larger than $\SI{600}{\milli\second}$ with our proposed solution as the best case.
The least complex thresholding method performs with an \ac{STO} in the units of seconds.

% -------------- Section end marker --------------
%                _       _
%               ( )_    ( )
%    ___  _   _ | ,_)   | |__     __   _ __   __
%  /'___)( ) ( )| |     |  _ `\ /'__`\( '__)/'__`\
% ( (___ | (_) || |_    | | | |(  ___/| |  (  ___/
% `\____)`\___/'`\__)   (_) (_)`\____)(_)  `\____)
%
% -------------- Section end marker --------------

% ============================================================
\section{Conclusion}
\label{sec_conclusion}
% ============================================================
\acresetall
This work introduced a physics-informed \ac{RL} framework for symbol synchronization in \ac{MC} systems.
The framework uses separate \ac{RC} architectures to implement the physics-informed module and an agent that learns the timing policy through a gradient-trained reservoir readout. 
Simulation results showed that the proposed framework closely preserves the physical likelihood landscape while reducing the \ac{STO} error compared with direct \ac{RC}-based timing inference.
Future work will pursue designs to synthesize the proposed framework on a chemical substrate and incorporate adaptive reservoir modules to operate under time-varying molecular channels.

\section*{Acknowledgement}
This work has been funded by the German Research Foundation (DFG) through the project SmartSynch, grant number TO 1422/4-1 and by the German Federal Ministry of Research, Technology and Space (BMFTR) through the project IoBNT, grant number 16KIS1986K.

\section{Appendix / Supplemental Material}
\label{app:details}
% ============================================================
\subsection{Deriving the echo state matrices from the molecular diffusion channel}
\label{sec_W_app}

We derive the matrices $\mathbf{W}$ and $\mathbf{W}_\mathrm{in}$ in \Cref{eq_reservoir} starting from the echo state definition for $\mathbf{x}[n]$ in \Cref{eq_x}.
The components of this vector are the echos of the input signal $u(t_n)$ to the reservoir, and within the diffusion channel in \Cref{fig_reservoir}~(a) it is produced by the binding of molecules that are released from a distance $d_\mathrm{RC}$.
We derive the echo-state matrices by finding the analytic relation between $x_m[n]$ and $u[n]$ in this diffusion channel.
From this relation, we identified the components of $\textbf{W}$ and $\textbf{W}_\mathrm{in}$ and later generalized to the matrix formulation.\footnote{Derivations in this appendix have been completed with the assistance of \ac{AI}.}

We depart from approximating the ligand-binding dynamics $db/dt=k_\mathrm{on}c(t)[1-b(t)]-k_\mathrm{off}b(t)$ with a discrete differential equation as,
\begin{equation}
\frac{db(t)}{dt}\approx k_\mathrm{on}c(t_n)\big[1-b(t_{n-1})\big]-k_\mathrm{off}b(t_{n-1}).
\end{equation}
Multiplying by $t_s$ and collecting the terms in $b(t_{n-1})$, yields
\begin{equation}\label{eq_euler}
b(t_n)=\big[1-t_s\big(k_\mathrm{off}+k_\mathrm{on}c(t_n)\big)\big]\,b(t_{n-1})
+t_sk_\mathrm{on}c(t_n),
\end{equation}
where we can identify with $c(t_n)$ their relation with the input $u(t_n)$ as $c(t_n)=\sum_{l=-\infty}^\infty u(t_{n-l})\,h(t_l),\text{with } t_l=l\,t_s$.

Because $c(t_n)$ in \Cref{eq_euler} appears twice and is multiplicative with the recurrent component $b(t_{n-1})$, this dependency does not allow us to express a linear relation for the echo state vector, as given in \Cref{eq_reservoir}.
We therefore evaluate the linearized version where we freeze the coefficient at the operating point $\bar c=\mathbb{E}[c(t)]$ and leaving the second term exact as
\begin{equation}\label{eq_frozen}
b(t_n)=\underbrace{\bar\alpha\,b(t_{n-1})+t_sk_\mathrm{on}c(t_n)}_{\text{linear component}}+\varepsilon_n, \text{ }
\bar\alpha=1-t_s\big(k_\mathrm{off}+k_\mathrm{on}\bar c\big),
\end{equation}
where $\varepsilon_n=-t_sk_\mathrm{on}\big(c(t_n)-\bar c\big)b(t_{n-1})$ is the discarded bilinear term, $\bar c$ is evaluated with their average over the time window $N$ as 
\begin{equation}\label{eq_bar_c}
\bar c=\frac{1}{N}\sum_{l=-N/2}^{N/2}{c(t_{n+l})},
\end{equation}
and 
\begin{equation}\label{eq_alpha}
    \bar{\alpha}=1-t_s\big(k_\mathrm{off}+k_\mathrm{on}\bar c\big).    
\end{equation}

We then proceed with the expansion of the linear component in \Cref{eq_frozen}, looking for its explicit relation between the vector components $[\mathbf{x}]_m$ and its dependency on the input $u(t_n)$.
The first component of $\mathbf{x}$, as given in \Cref{eq_x}, is $x_1[n]=b(t_n)$ and the remaining components are pure delays versions of the previous components as $x_{m}[n]=x_{m-1}[n-1]$.
Hence making the contribution of every entry of $\mathbf{x}[n]$ visible we have the following,
\begin{equation}\label{eq_rows}
\begin{aligned}
x_1[n]&=\underbrace{\bar\alpha\,x_1[n-1]+0\cdot x_2[n-1]+\cdots+0\cdot x_N[n-1]}_{\text{recursive component}}\\
&+\underbrace{t_sk_\mathrm{on}\!\!\sum_{l=0}^{N L_\mathrm{RC}-1}\!\!h(t_l)\,u[n-l]}_{\text{input component}},\\
x_2[n]&=1\cdot x_1[n-1]+0\cdot x_2[n-1]+\cdots+0\cdot x_N[n-1],\\
%x_3[n]&=0\cdot x_1[n-1]+1\cdot x_2[n-1]+\cdots+0\cdot x_N[n-1],\\
&\;\;\vdots\\
x_N[n]&=0\cdot x_1[n-1]+\cdots+1\cdot x_{N-1}[n-1]+0\cdot x_N[n-1],
\end{aligned}
\end{equation}
where each line represents one row of the recursion, with the coefficients multiplying $x_1[n-1],\dots,x_N[n-1]$ forming the corresponding row of $\mathbf{W}$.
After stacking these lines, we get the recursive matrix for $\mathbf{W}$ in \Cref{eq_matrices}.

To derive $\textbf{W}_\mathrm{in}$, we note that the input $u(t_n)$ feeds only the first node of the state vector $\textbf{x}[n]$ in~\Cref{eq_x} via the diffusion and ligand receptor process.
That is, the relation between the input and the state vector follows the chain $u(t_n)\rightarrow c(t_n)\rightarrow x_1[n]$, after completing the update $\mathbf{x}_{N-1}[n-1]\rightarrow\mathbf{x}_{N}[n]\,\dots,\mathbf{x}_1[n-1]\rightarrow\mathbf{x}_2[n]$. 
Therefore, as $\textbf{x}[n]$ is defined as a delay line, only the first row of $\mathbf{W}_{\mathrm{in}}$ is nonzero.
Besides, to formulate a linear dependency of $\mathrm{x}[n]$ with $\mathrm{u}[n]$ components via the diffusion process in the molecular channel, we use a truncated version of $c(t_n)$.
We restrict the convolution to the finite channel memory $L_{\mathrm{RC}}$ as
\begin{equation}\label{eq_c_app}
    c(t_n)\approx\sum_{l=0}^{N\!L_{\mathrm{RC}}-1}u(t_{n-l})h(t_{l}),
\end{equation}
which also accounts for the causility of the \ac{CIR}, i.e., $h(t)=0$ when $t<0$.
Consequently, only inputs from the previous $L_{\mathrm{RC}}$ symbol intervals, corresponding to $NL_{\mathrm{RC}}$ discrete samples, contribute to the current concentration $c(t_n)$.
Stacking the $N$ lines that expand from $t_{n-N\!L_\mathrm{RC}}$ to $t_n$ yields the form for $\mathbf{W}_\mathrm{in}{\mathbf{u}}[n]$, with $\mathbf{W}_\mathrm{in}$ as given in \Cref{eq_matrices}.

The matrix $\mathbf{W}$ is lower bidiagonal, where $\bar\alpha$ acts on the most recent occupancy sample, and the unit subdiagonal shifts the window, i.e., represents the delay line in \Cref{eq_x}.
Only the first row of $\mathbf{W}_\mathrm{in}$ is populated, since the diffusion channel drives a single receptor population.
In this case, the \ac{CIR} therefore plays the role of the input mask, in contrast with the random projection of a conventional echo state network~\cite[Eq.~(1)]{jaeger2001echo}.

\subsection{Echo state property}
\label{sec_echo_app}

The echo state property requires that the influence of the initial condition $\mathbf{x}[0]$ on the echo vector vanishes asymptotically, so that $\mathbf{x}[n]$ becomes a function of the input history alone \cite[Def.~1]{jaeger2001echo}.
As a consequence, for every input sequence $u\in U^{\mathbb{Z}}$ and every pair of initial states $\mathbf{x}[0],\mathbf{x}'[0]\in A$ driven by that same input, the condition
\begin{equation}
  \left| x_m[n]-x'_m[n] \right| \xrightarrow[n\to\infty]{} 0,
  \qquad m=1,\dots,N,
\end{equation}
most hold, where $U=[0,M]$ is the set of admissible molecule counts released by the point transmitter and $A=\{\mathbf{x}\in\mathbb{R}^N: 0\le x_m\le 1\}$, since each component in $\mathbf{x}[n]$ is a fraction of occupied receptors in the range $\{0,\,1\}$.
Because $x_m[n]=b(t_{n-m+1})$, it suffices to establish the property for the receptor recursion itself.
Unrolling \Cref{eq_euler} to $t_0$, and without invoking the linearisation of \Cref{eq_euler}, yields
\begin{equation}
  \left| b(t_n) - b'(t_n) \right|
  = \prod_{j=1}^{n}\left| 1 - t_s\big(k_{\mathrm{off}}+k_{\mathrm{on}}c(t_j)\big)\right|
    \cdot \left| b(t_0)-b'(t_0) \right|.
  \label{eq_contraction}
\end{equation}
Thus, a sufficient condition for the echo state property to hold is whenever the factors are uniformly bounded below unity,
\begin{equation}\label{eq_sup_cond}
\sup_{j}\text{ } t_s\big(k_{\mathrm{off}}+k_{\mathrm{on}}c(t_j)\big) < 1.    \end{equation}

%The same condition keeps every factor positive, which guarantees $b\in[0,1]$ and hence the invariance of $A$ under the update.

Since $c(t_n)=\sum_{l=-\infty}^\infty u(t_{n-l})h(t_l)$ with $h\ge 0$, the supremum is attained by the largest admissible release superposed over the channel memory, and is bounded in closed form by %$c_{\max} \le M\sum_{l\ge0} h(t_l)\approx \frac{M}{t_s}\int_0^{\infty} h(t)\,dt = \frac{M}{4\pi D\,d_{\mathrm{RC}}\,t_s}$,
\begin{equation}\label{eq_c_max}
  c_{\max} \le M\sum_{l\ge0} h(t_l)
  \approx \frac{M}{t_s}\int_0^{\infty} h(t)\,dt
  = \frac{M}{4\pi D\,d_{\mathrm{RC}}\,t_s},
\end{equation}
where $h(t)=\frac{1}{(4\pi D t)^{3/2}}e^{\frac{-d_\mathrm{RC}^2}{4Dt}}$ as follows from \cite[Eq. (4)]{uzun2026molecular}, and after using the integral result $\int_0^\infty t^{-3/2}e^{-a/t}dt=\pi/a$.
Substituting  $c_{max}$ into \Cref{eq_sup_cond} yields a sufficient condition for the molecular channel reservoir to fulfill the echo state property, expressed directly in the design parameters for $d_\mathrm{RC}$ in \Cref{eq_cond}.

We also derive the sufficient condition for the echo-state property with the linear model in \Cref{eq_matrices}, where the maximum eigenvalue of $\textbf{W}$ is less than one, i.e., $\rho(\textbf{W})<1$~\cite[Sec. 27.3.2]{lukosevicius2012practical}.
To find $\rho(\textbf{W})$, we find the eigenvalues of the matrix $\textbf{W}$ using the definition straigthforwardly.
Subtracting $\lambda$ from the diagonal of $\mathbf{W}$, given in \Cref{eq_matrices}, yields a lower triangular matrix.
Since the determinant of a triangular matrix is the product of its diagonal entries, the determinant is readily given by
\begin{equation}\label{eq_charpoly}
\det(\mathbf{W}-\lambda\mathbf{I})
=(\bar\alpha-\lambda)\prod_{m=2}^{N}(-\lambda)=(\bar\alpha-\lambda)(-\lambda)^{N-1},
\end{equation}
whose solutions are $\lambda=\bar\alpha$ and $\lambda=0$ with multiplicity
$N-1$, thus the eigenvalues are $\{\bar\alpha,0,\dots,0\}$.
The delay line constructed by the reservoir nodes contributes only null
modes, and the single non-zero eigenvalue is the receptor leak.
The spectral radius then yields the equation illustrated in \Cref{eq_rho}.
The echo state necessary condition $\rho(\mathbf{W})<1$ holds whenever
\begin{equation}\label{eq_esp_cond}
0<t_s\big(k_\mathrm{off}+k_\mathrm{on}\bar c\big)<2.
\end{equation}
The left inequality is satisfied by any positive unbinding rate and the right one can be fulfilled with the value of $\bar{c}$ (\Cref{eq_bar_c}) which is ultimately defined by $d_\mathrm{RC}$.

\printbibliography
\end{document}